\documentclass[%
 aip,
 amsmath,amssymb,
 reprint,%
nofootinbib    
]{revtex4-1}

\usepackage{graphicx}
\usepackage{dcolumn}
\usepackage{bm}
\usepackage{caption}    
\usepackage{subcaption}    
\usepackage{array}  
\usepackage[hidelinks,pagebackref]{hyperref}	
\usepackage{xcolor}   
\usepackage{soul}   
\usepackage[export]{adjustbox}  
\usepackage{gensymb}    

\usepackage{titlesec}
\titlespacing*{\section} {0pt}{0.8ex}{0.2ex}    

\usepackage[utf8]{inputenc}
\usepackage[T1]{fontenc}
\usepackage{mathptmx}
\usepackage{etoolbox}
\usepackage{environ}         

\DeclareUnicodeCharacter{2060}{!!!!I-AM-HERE!!!!!}   
\DeclareUnicodeCharacter{03BD}{!!!!I-AM-HERE!!!!!}   
\DeclareUnicodeCharacter{03C3}{!!!!I-AM-HERE!!!!!}   

\makeatletter
\def\@email#1#2{%
 \endgroup
 \patchcmd{\titleblock@produce}
  {\frontmatter@RRAPformat}
  {\frontmatter@RRAPformat{\produce@RRAP{*#1\href{mailto:#2}{#2}}}\frontmatter@RRAPformat}
  {}{}
}%
\makeatother

\newlength{\myl}
\let\origequation=\equation
\let\origendequation=\endequation

\RenewEnviron{equation}{
  \settowidth{\myl}{$\BODY$}                       
  \origequation
  \ifdimcomp{\the\linewidth}{>}{\the\myl}
  {\ensuremath{\BODY}}                             
  {\resizebox{\linewidth}{!}{\ensuremath{\BODY}}}  
  \origendequation
}

\begin{document}

\title{Evolution of radiation profiles in a strongly baffled divertor on MAST Upgrade
}

\author{Fabio Federici}

\altaffiliation{E-mail address: fabio.federici@ukaea.uk}
\affiliation{ 
Oak Ridge National Laboratory, Oak Ridge, Tennessee 37831, USA
}%
\author{Matthew L. Reinke}%
\affiliation{ 
Commonwealth Fusion Systems, Cambridge, MA 02139, USA
}%

\author{Bruce Lipschultz}%
\affiliation{ 
York Plasma Institute, University of York, Heslington, York, YO10 5DD, UK
}%

\author{Jack J. Lovell}%
\affiliation{ 
Oak Ridge National Laboratory, Oak Ridge, Tennessee 37831, USA
}%

\author{Kevin Verhaegh}%
\affiliation{ 
UK Atomic Energy Authority, Culham Centre for Fusion Energy, Abingdon, OX14 3DB, UK
}%

\author{Cyd Cowley}%
\affiliation{ 
UK Atomic Energy Authority, Culham Centre for Fusion Energy, Abingdon, OX14 3DB, UK
}%
\affiliation{ 
digiLab, The Quay, Exeter EX2 4AN, United Kingdom 
}%

\author{Mike Kryjak}%
\affiliation{ 
UK Atomic Energy Authority, Culham Centre for Fusion Energy, Abingdon, OX14 3DB, UK
}%
\affiliation{ 
York Plasma Institute, University of York, Heslington, York, YO10 5DD, UK
}%

\author{Peter Ryan}%
\affiliation{ 
UK Atomic Energy Authority, Culham Centre for Fusion Energy, Abingdon, OX14 3DB, UK
}%

\author{Andrew J. Thornton}%
\affiliation{ 
UK Atomic Energy Authority, Culham Centre for Fusion Energy, Abingdon, OX14 3DB, UK
}%

\author{James R. Harrison}%
\affiliation{ 
UK Atomic Energy Authority, Culham Centre for Fusion Energy, Abingdon, OX14 3DB, UK
}%




\author{Byron J. Peterson}%
\affiliation{ 
National Institute for Fusion Science, 322-6 Oroshi-cho, Toki 509-5292, Japan
}%

\author{Bartosz Lomanowski}%
\affiliation{ 
Oak Ridge National Laboratory, Oak Ridge, Tennessee 37831, USA
}%

\author{Jeremy D. Lore}%
\affiliation{ 
Oak Ridge National Laboratory, Oak Ridge, Tennessee 37831, USA
}%

\author{Yacopo Damizia}%
\affiliation{ 
UK Atomic Energy Authority, Culham Centre for Fusion Energy, Abingdon, OX14 3DB, UK
}%
\affiliation{ 
Electrical Engineering and Electronics department, University of Liverpool,Liverpool, L69 3GJ, UK
}%

\author{the EUROfusion Tokamak Exploitation Team}%
\altaffiliation{ 
See author list of E. Joffrin et al 2024 Nucl. Fusion (https://doi.org/10.1088/1741-4326/ad2be4)
}%

\author{the MAST Upgrade team}%
\altaffiliation{ 
See author list of J. Harrison et al 2019 Nucl. Fusion 59 112011 (https://doi.org/10.1088/1741-4326/ab121c)
}%


\begin{abstract}

\fontsize{7.5pt}{8pt}\selectfont

Plasma detachment in tokamaks is useful for reducing heat flux to the target. It involves interactions of the plasma with impurities and neutral particles, leading to significant losses of plasma power, momentum, and particles. Accurate mapping of plasma emissivity in the divertor and X-point region is essential for assessing the relationship between particle flux and radiative detachment. The recently validated InfraRed Video Bolometer (IRVB) diagnostic, in MAST-U\cite{Federici2023} enables this mapping with higher spatial resolution than more established methods like resistive bolometers.

In previous preliminary work\cite{Federici2023a}, the evolution of radiative detachment was characterised in L-mode (power entering the scrape-off layer, $P_{SOL} \sim $0.4MW). With a conventional divertor the inner leg consistently detached ahead of the outer leg, and radiative detachment preceded particle flux detachment. 
This work presents results also from the third MAST-U experimental campaign, fuelled from the low field side instead of the high field side, including Ohmic and beam heated L-mode shots (with a power exiting the core up to $P_{SOL} \sim$ 1-1.5MW).

The radiation peak moves upstream from the target at lower upstream densities than the ion target flux roll-over (typically considered the detachment onset), while the inner leg detaches before the outer one. The movement of the radiation is in partial agreement with the expectations from the DLS model\cite{Myatra2021,Cowley2022,Lipschultz2016}, predicting a sudden shift from the target to the X-point. The energy confinement is found to be related to detachment, but there seems to be some margin between the radiation on the inner leg reaching the X-point and confinement being affected, a beneficial characteristic if it could be extrapolated to future reactors. For increasing $P_{SOL}$ the particle flux roll over is almost unaffected, while radiative detachment occurs at higher density in both legs, but much higher on the outer, suggesting an uneven distribution of the power exiting the core.

\end{abstract}

\maketitle


\section{Introduction}\label{introduction}

MAST-U is a spherical tokamak at the Culham Centre for Fusion Energy (CCFE) in the United Kingdom\cite{Morris2018,
Fishpool2013}. It features a double null (DN) plasma, strongly baffled divertor configurations, and can support an innovative Super-X divertor (SXD), which significantly reduces target heat loads and improves access and stability of plasma (e.g., 'particle') detachment\cite{Moulton2024,Verhaegh2022,Verhaegh2023a}. 

In this work, we investigate the radiative power dissipation and its evolution as detachment progresses on MAST-U. To accurately measure the total emissivity profile, multiple resistive bolometer arrays are installed to monitor the core and divertor chamber. To complement the resistive system and fill the gap from the X-point to the divertor chamber (see \autoref{fig:res_bolo1}), a prototype Infrared Video Bolometer (IRVB) was installed, aimed at the lower X-point. This diagnostic was recently validated\cite{Federici2023} and its data have already been used to complement various scientific endeavors\cite{Soukhanovskii2022a,Verhaegh2023a,Moulton2024,Henderson2024}. Until now, the resistive system has been affected by significant noise that, while still allowing the calculation of the total radiated power from the core, prevented detailed measurements of the movement of the radiation front in the divertor chamber. Although the IRVB cannot reconstruct the radiative emissivity map in the divertor chamber downstream the baffle, due to its viewing location, its fiend of view (FOV) is adequate for investigating the radiative emissivity distribution in the Conventional Divertor (CD) configuration\cite{Federici2023b}.

This paper presents the initial results from the analysis of the IRVB data from the first MAST-U experimental campaign (MU01) and third (MU03) L-mode CD shots focusing on changes in the total radiated power spatial distribution along the divertor legs in connection with detachment.

Usually a region with high total radiative emissivity is present on the divertor legs at the strikes points or near scrape-off layer (SOL). When the core density increases, this region (also called the radiation front) moves upstream towards the X-point, 'detaches' from the target, moving along the separatrix. This is related with the ionisation front detachment and was the subject of significant modeling efforts using simplified analytical models, among which is the detachment location sensitivity (DLS) model, which aims to predict the location and sensitivity of the front\cite{Myatra2021,Cowley2022,Lipschultz2016} and will be here compared with experimental observations.


\section{Diagnostic improvements}\label{Experimental setup}

\begin{figure}
     \centering
     \begin{subfigure}{0.38\linewidth}
         \centering
         \includegraphics[trim={460 30 130 40},clip,width=\textwidth]{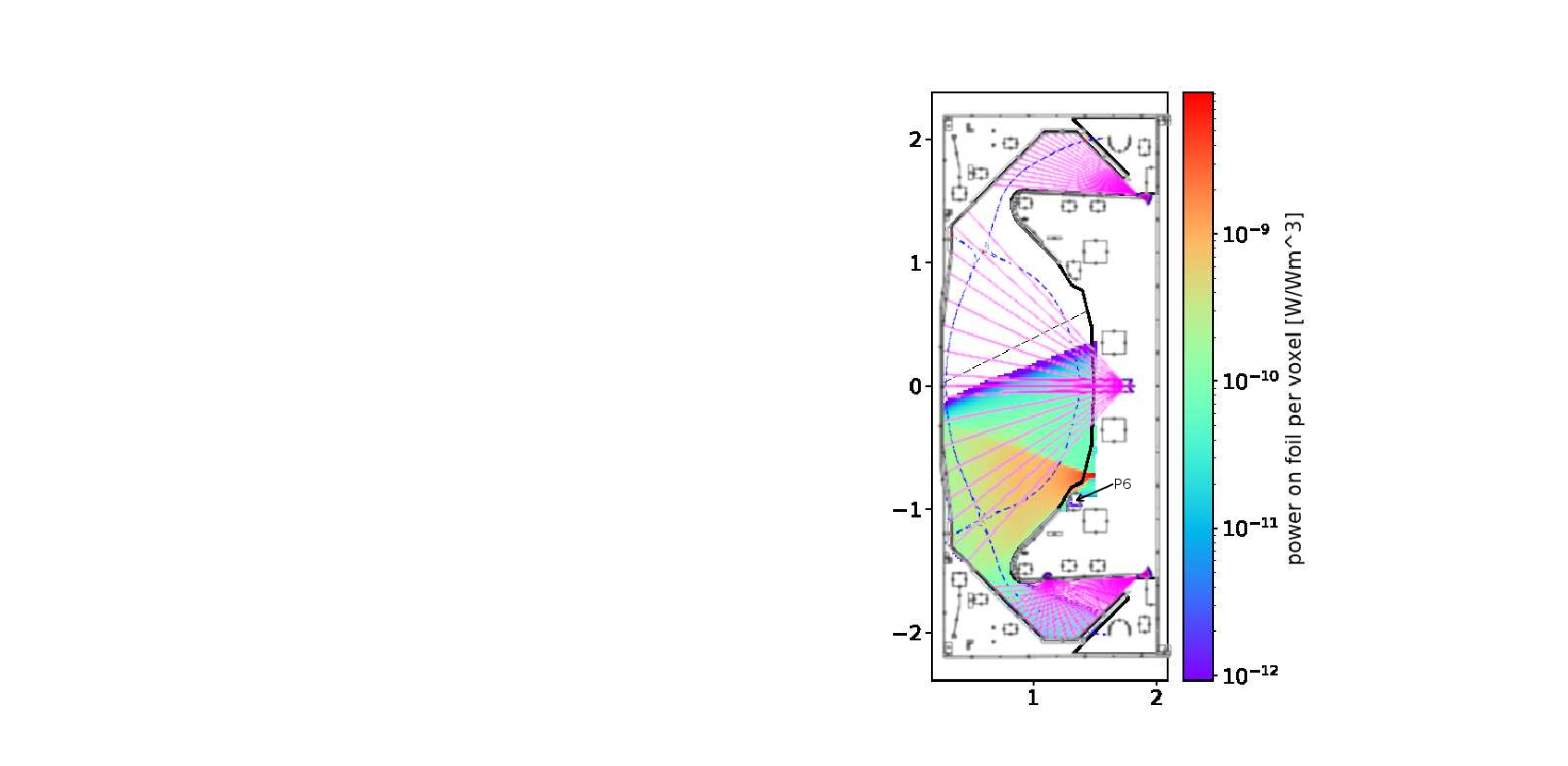}
        \vspace{-7mm}
         \caption{}
         \label{fig:res_bolo1a}
     \end{subfigure}
     \begin{subfigure}{0.45\linewidth}
         \centering
         \includegraphics[trim={109 160 70 10},clip,width=\textwidth]{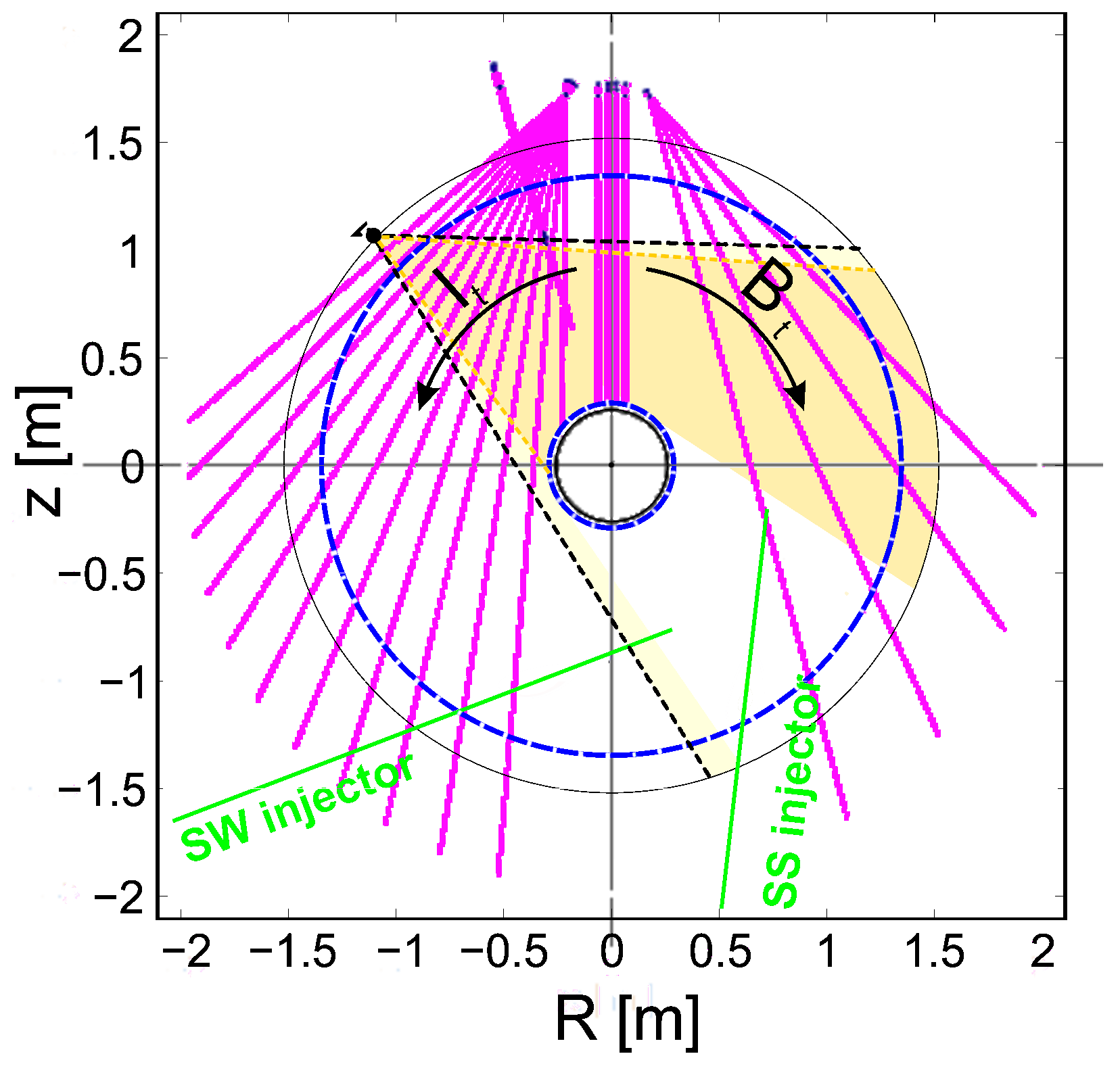}
    \vspace{-3mm}
    \caption{}
         \label{fig:res_bolo1b}
     \end{subfigure}
    \vspace{-3mm}
    \caption{Poloidal (\subref{fig:res_bolo1a}) and top (\subref{fig:res_bolo1b}) view of MAST-U, showing the comparison of the resistive bolometer system LOS (magenta) with (\subref{fig:res_bolo1a}) a color plot obtained by scanning all the voxels with a $1W/m^3$ emitter and integrating the power absorbed by the foil, indicating the regions of higher sensitivity of the IRVB, and (\subref{fig:res_bolo1b}) the edges of the FOV (yellow, mostly counter-NBI). In  (\subref{fig:res_bolo1b}) is also the position of neutral beam injectors (NBI, green). Adapted from \cite{Rivero-Rodriguez2018}. For reference the separatrix of a typical plasma is shown as an overlay of a blue dashed line. The FOV for MU02 and MU03 is shown and the edges of the MU01 FOV are shown with dashed black lines.}
    \label{fig:res_bolo1}
    \vspace*{-7mm}
\end{figure}

Before discussing our experimental results, we will first discuss the IRVB diagnostic implementation and its various improvements. IRVB measurements first started in MU01, further documented in references \cite{Federici2023,Federici2023b}. 
The geometry of the IRVB was optimised in MU02 to provide a more detailed view of the plasma around the X-point by retracting the foil from the pinhole, from 45mm to 60mm. A significantly thinner platinum absorber foil than expected (measured $\sim 0.72 \mu m$ instead of nominal $2.5$)\cite{Federici2024} resulted in higher signal levels than expected \cite{Federici2023}, enabling this modification. The IRVB geometry optimisation also reduced the portion of the foil shaded by the P6 coil (see \autoref{fig:res_bolo1}) and increased the coverage of the divertor chamber (Figure 2.5a to 2.5b in \cite{Federici2023b}). The IRVB geometry was verified to improve the accuracy of the geometrical calibration beyond the design specification. The internal pinhole location was accurately triangulated with sub-mm precision with CALCAM fits from multiple angles\cite{Silburn2020}, returning a $\sim$3.9mm shift with respect to target parameter. Together with the exact location of the IRVB flange on the vacuum vessel, the precise FOV for MU02, MU03, and the future MU04 was determined, shown in \autoref{fig:res_bolo1}. This improved IRVB FOV characterisation has greatly improved the accuracy of IRVB measurements, enabling it to distinguish radiation at or slightly (a few cm) away from the plasma surface facing components in the divertor.

After a camera image is obtained by the IRVB, a Bayesian tomographic inversion is performed to obtain a 2D radiative emissivity map. This inversion is performed with an arbitrary regularisation coefficient to reduce noise on the inversion. Unlike previous results, spatial binning has been disabled, improving the inversion by making use of the full resolution available from the camera. A running average smoother ($\sim$ 30ms) is applied over time to remove temporal oscillations presented previously\cite{Federici2023}.

\section{Radiative emissivity results}\label{Radiative emissivity results}

L-mode DN shots are analysed in this paper, as upstream conditions are difficult to control in H-mode. The data is from the MU01 and MU03 campaigns. The only impurity present in significant quantities is the intrinsic carbon from the walls.

MU01 was the first experimental campaign in MAST-U and it was often characterised by the presence of MHD activity (possibly influenced by error fields) and imprecise plasma control, which negatively affected the overall plasma performance (shots 45468, 45469, 45470, 45473). The shots are Ohmically heated with fueling from the high field side (HFS), have a plasma current $I_p=600kA$ and have a power crossing the separatrix ($P_{SOL}$) $\sim 0.4MW$ (determined as the Ohmic power minus the power radiated in the core from resistive bolometry).

\begin{figure}
     \centering
     \begin{subfigure}{0.35\linewidth}
         \centering
         \includegraphics[trim={30 140 10 178},clip,width=\textwidth]{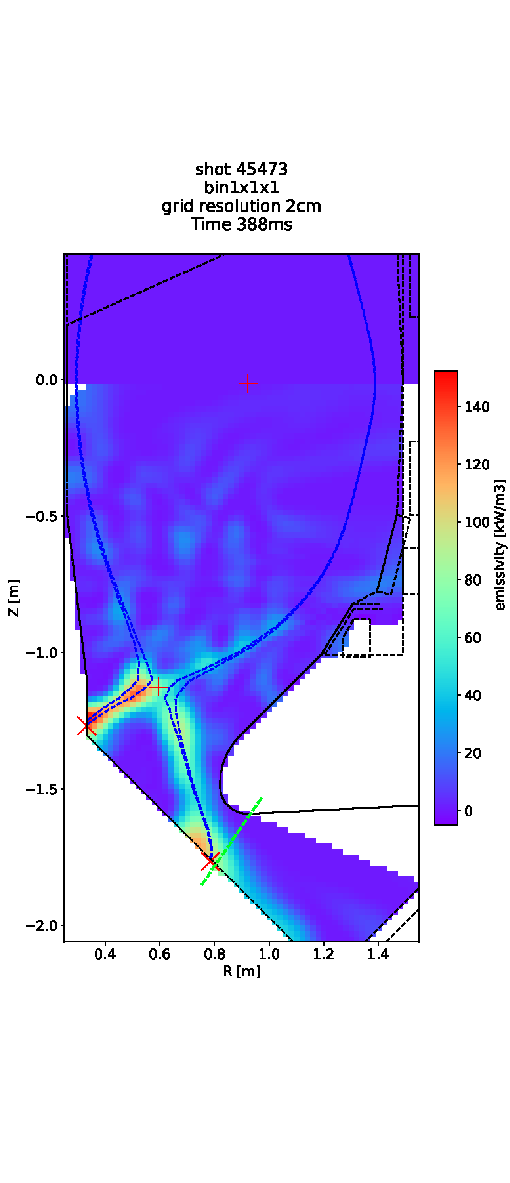}
         \vspace*{-4.5mm}
         \caption{388ms, $n_u=0.25 \times 10^{19} \#/m^3$}
         \label{fig:45473_export2_1}
     \end{subfigure}
     \begin{subfigure}{0.35\linewidth}
         \centering
         \includegraphics[trim={30 140 10 178},clip,width=\textwidth]{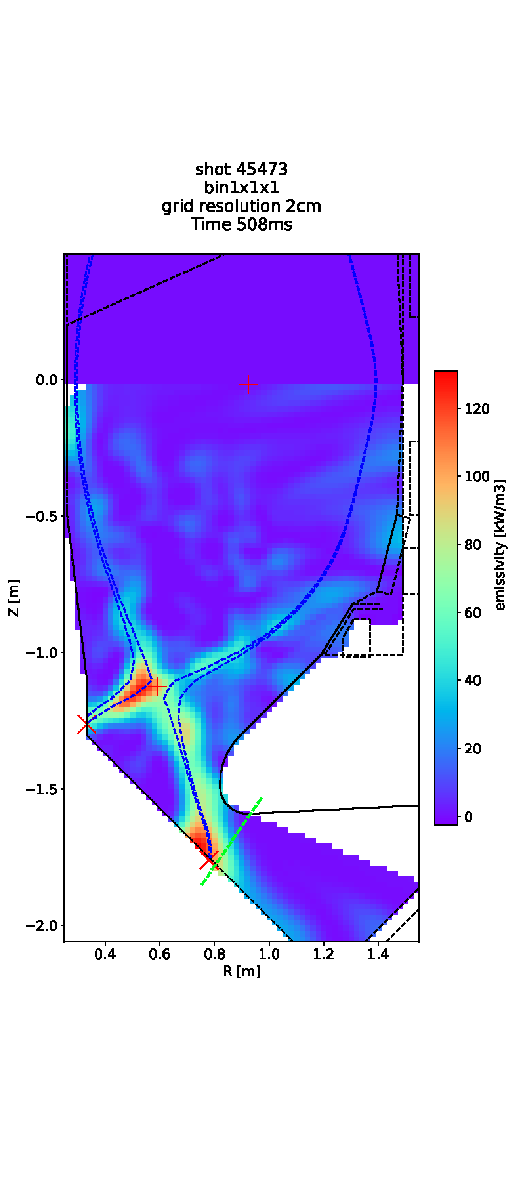}
         \vspace*{-4.5mm}
         \caption{508ms, $n_u=0.34 \times 10^{19} \#/m^3$}
         \label{fig:45473_export2_2}
     \end{subfigure}
     \begin{subfigure}{0.35\linewidth}
         \centering
         \includegraphics[trim={30 140 10 178},clip,width=\textwidth]{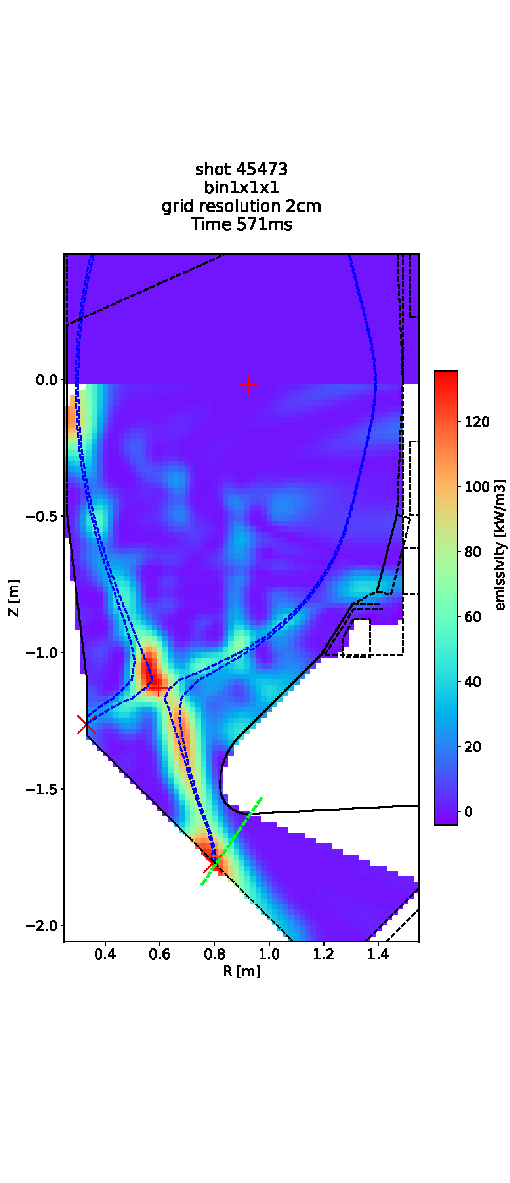}
         \vspace*{-4.5mm}
         \caption{571ms, $n_u=0.46 \times 10^{19} \#/m^3$}
         \label{fig:45473_export2_3}
     \end{subfigure}
     \begin{subfigure}{0.35\linewidth}
         \centering
         \includegraphics[trim={30 140 10 178},clip,width=\textwidth]{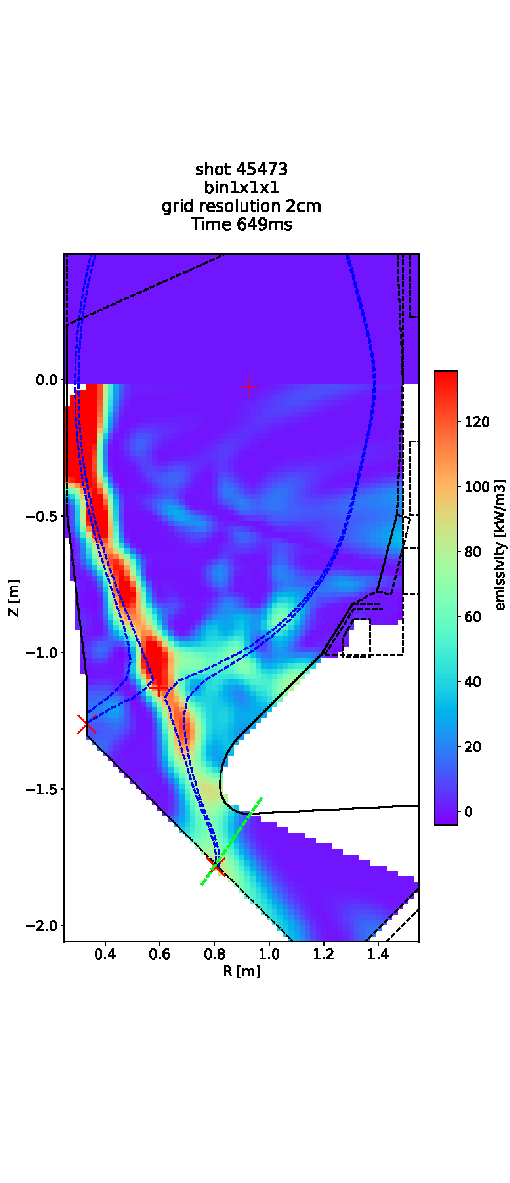}
         \vspace*{-4.5mm}
         \caption{649ms, $n_u=0.54 \times 10^{19} \#/m^3$}
         \label{fig:45473_export2_4}
     \end{subfigure}
     \begin{subfigure}{0.35\linewidth}
         \centering
         \includegraphics[trim={30 140 10 178},clip,width=\textwidth]{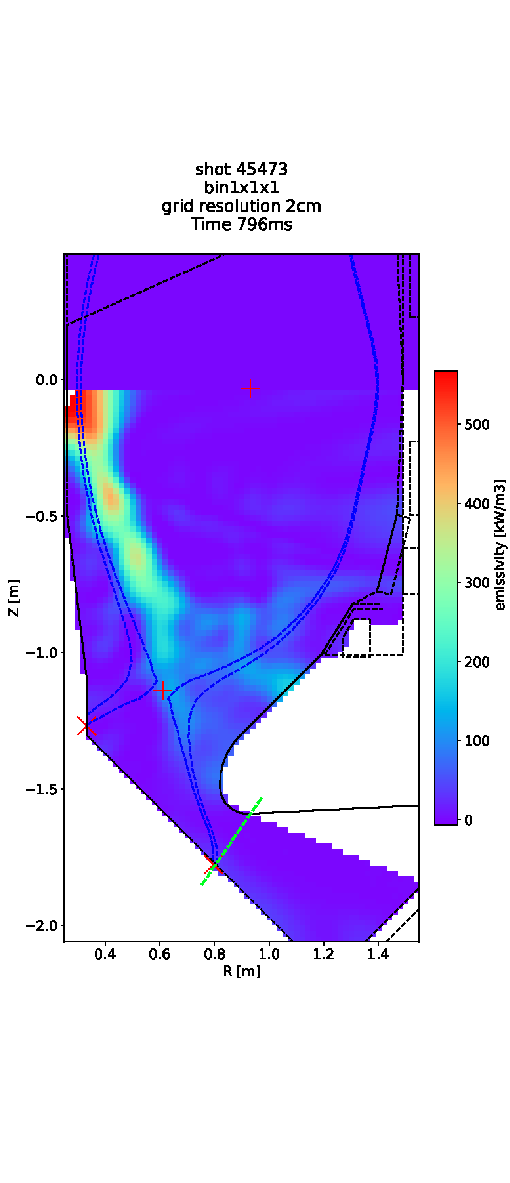}
         \vspace*{-4.5mm}
         \caption{796ms, $n_u=0.63 \times 10^{19} \#/m^3$}
         \label{fig:45473_export2_5}
     \end{subfigure}
     \begin{subfigure}{0.03\linewidth}
         \centering
         \includegraphics[trim={236 240 0 220},clip,width=\textwidth]{figs/IRVB-MASTU_shot-45473_export_81.eps}
     \end{subfigure}
    \vspace*{-3mm}
    \caption{Emissivity distribution in a density ramp for a conventional divertor, L-mode, Ohmic plasma (shot 45473, DN and $I_p=600kA$, same used in \cite{Federici2023}). First, both inner and outer targets are radiatively attached (\subref{fig:45473_export2_1}), then the inner detaches (\subref{fig:45473_export2_2}) up to the X-point (\subref{fig:45473_export2_3}), and finally the outer leg detaches and a radiation MARFE-like structure grows on the high field side (HFS) midplane (\subref{fig:45473_export2_4}). Further increasing the density, the structure moves inward (\subref{fig:45473_export2_5}) leading to a disruption. Note that all images have different color bar ranges. The limit below which the radiation distribution is considered unreliable as per \cite{Federici2023b} is marked in dashed green.}
    \label{fig:45473_export2}
    \vspace*{-8mm}
\end{figure}

In MU03 similar shots were performed with a more optimised scenario, yielding better overall performance, but with a higher starting density, so the transition of the radiation detaching on both legs cannot be observed (shots 47950, 47973, 48144). The shots are Ohmically heated with main fueling from the low field side (LFS), $I_p=750kA$ and $P_{SOL} \sim 0.5-0.6MW$.

LFS fuelling was employed to enable higher power L-mode operation, making the scenario compatible with off-axis (SW) 1.5-1.8 MW NBI heating \cite{Verhaegh2023b} by raising the L-H threshold. These beam-heated L-mode discharges allow us to verify if the detachment evolution changes with a higher $P_{SOL}$ and to better probe the initial stages of detachment (at higher $q_\parallel$ a higher $n_u$ is expected to be required to achieve detachment), featuring $I_p=750kA$ and $P_{SOL} \sim 1-1.5MW$.

First we will describe the typical evolution of the radiative emissivity during a core density ramp where the divertor is progressively cooled in
\autoref{fig:45473_export2}. This indicates that the radiation region detaches from the inner target at lower densities than the outer target. The outer leg dissipates significantly more power than the inner one, due to its larger volume integral. After the inner leg radiation moves towards the X-point, it progresses upstream the X-point along the inner separatrix before the target radiation disappears from the outer strike point. A MARFE-like structure appears at the HFS midplane even before the outer leg radiatively detaches.  This is likely caused by HFS fuelling, as the inner gap ($\sim 4$ cm) is sufficiently large to avoid interactions with the HFS column and this phenomenon does not occur in shots fuelled from the LFS.
The presence of the MARFE-like structure is observed in interpretative SOLPS simulations \cite{Moulton2024} and is confirmed by high speed visible light imaging, Thomson scattering (TS) measurements (that shows a region of high density and low temperature penetrating the core from the HFS midplane) and resistive bolometry (increased brightness of the LOS close to the central column). The latter is routinely used to verify if a peaked emission towards the inner midplane in the IRVB inversions constitutes an artefact or a true MARFE-structure occurrence (phenomena more common with the optimised IRVB FOV).

Although the spatial resolution is insufficient to distinguish core and SOL radiation near the separatrix, a clear inward movement towards the core is observed at higher electron densities, by comparing  \autoref{fig:45473_export2_3} and \ref{fig:45473_export2_4} with\autoref{fig:45473_export2_5}). 

\section{Model predictions}\label{Model predictions}

Our aim is to compare the movement of the total radiation distribution against model expectations. The Two Point Model (2PM)\cite{
Stangeby2001} was later modified to include the presence of a thermal or radiation front in the Thermal Front Model\cite{Hutchinson1994}. This model was further refined to consider the leg geometry and magnetic field, resulting in the Detachment Location Sensitivity (DLS) model\cite{Lipschultz2016}. 
Assuming that thermal front movement corresponds to detachment, this reduced model predicts that for given magnetic geometry (e.g., different inner and outer leg topologies (CD, Super-X)) the location of the detachment front depends on the control parameter $C = \frac{n_u \sqrt{f_I}}{q_{\parallel,u}^{5/7}}$, containing upstream conditions (upstream heat flux $q_{\parallel, u}$, upstream electron density $n_u$) and the divertor impurity concentration $f_I$. The impact of the magnetic geometry on the detachment onset and evolution can be quantified via the coefficient $C_1 (s_{\parallel, f}) = \frac{{B_f}}{B_u} \left[ \int_{s_{X}}^{L_{\parallel}} \frac{B(s_{\parallel})}{B_X} \frac{(L_{\parallel} - s_{\parallel})}{(L_{\parallel} - s_{X})} \,ds_{\parallel} + \int_{s_{\parallel,f}}^{s_{X}} \frac{B(s_{\parallel})}{B_X} \,ds_{\parallel} \right]^{-2/7}$ where $s_{\parallel}$ \cite{Cowley2022} is the position along the separatrix and $f$ denotes quantities at the front location ($S_{\parallel, f} = 0$ corresponds to the detachment front at the target: the detachment onset). Assuming a constant electron heat conductivity coefficient and radiative cooling function (it depends on the impurity specie and its transport), $\frac{n_u \sqrt{f_I}}{q_{\parallel,u}^{5/7}} \propto C_1$. Therefore, the front location can be modelled as function of magnetic geometry $C_1 (s_{\parallel, f})$ in terms of changes in $n_u$, $f_I$, $q_{\parallel,u}$. If $f_I$ and $P_{SOL}$ are constant and the MAST scaling law between power decay length at the outer midplane ($\lambda_q$) and $n_u$ from \cite{Harrison2013} is used, this further reduces to ${n_u}^1.5 \propto C_1 (s_{\parallel, f})$. The sensitivity and stability of the detachment front at a certain location ($s_{\parallel,f}$) can be modelled as $\frac{d C_1}{d s_{\parallel, f}}$. If $C_1$ decreases from the target to the X-point ($\frac{d C_1}{d s_{\parallel, f}} < 0$), the detachment front location is unstable, as a slight perturbation towards the X-point increases the power dissipated in it, pushing the front further upstream. Conversely, $\frac{d C_1}{d s_{\parallel, f}} > 0$ implies an intrinsic detachment front stabilisation. 

\begin{figure}
    \centering
    \includegraphics[trim={0 0 0 35},clip,width=0.75\linewidth]{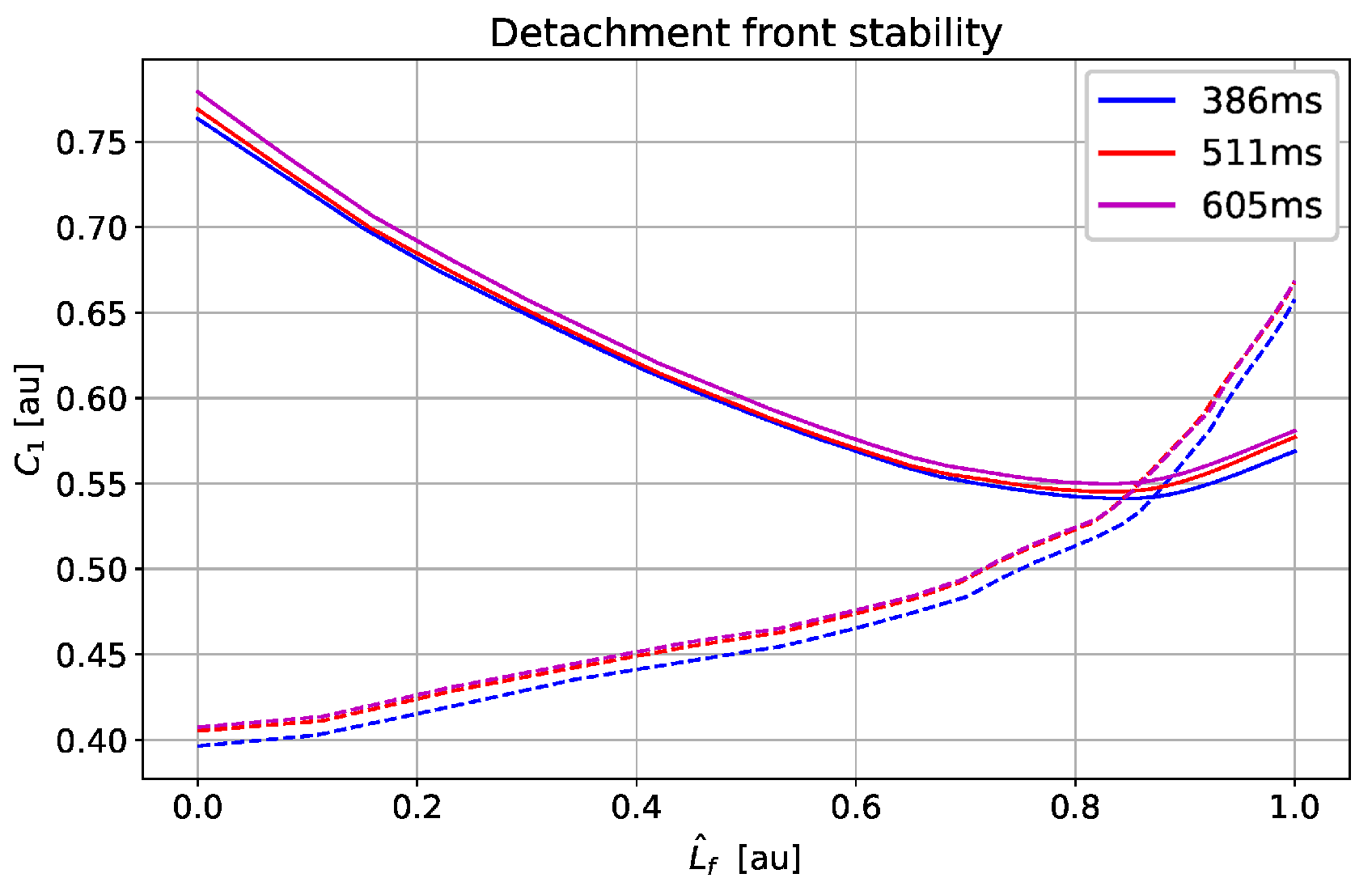}
    \vspace{-3mm}
    \caption{$C_1$ parameter variation in shot 45473 for a front position from target ($\hat{L}_f$=0) to X-point ($\hat{L}_f$=1). Solid lines for inner leg, dashed for outer.\cite{Federici2023b}}
    \label{fig:DLS}
    \vspace*{-8mm}
\end{figure}

For a typical MAST-U CD configuration, the $C_1$ profile along the inner and outer divertor leg is shown in \autoref{fig:DLS}.
The front on the inner leg is unstable up to a location very close to the X-point, while the front on the outer leg is stable. This model behaviour can be verified with IRVB measurements.

\section{Front characterisation}\label{Front characterisation}

To compare the DLS predictions with the IRVB data, it is essential to define what is meant with the term (detachment) \emph{front}. In the DLS and other models, this is defined as an infinitely narrow region where "\emph{the electron temperature transitions between the hotter upstream region and the colder region below which is dominated by ionization, recombination, and other neutral processes}"\cite{Lipschultz2016}. This region is associated with high total radiated power, usually attributed to the presence of impurities in the plasma that radiate efficiently at the temperature typical of the SOL and divertor, causing the necessary power losses. The front is idealised in the DLS model as infinitely narrow, so the (impurity) radiation front and the ionisation (detachment) front coincide. In reality this is not the case and a clear separation between the impurity radiation region and the ionisation/detachment front is observed \cite{Verhaegh2023b}. Furthermore, the emissivity seldom has a single well defined peak that identifies the radiation front. 

A simple way is to identify the front as the region with the peak emissivity along a divertor leg and thus the highest radiative dissipation. The current MAST-U IRVB implementation, though, is characterised by significant absorber foil properties non-uniformities\cite{Federici2023}, which causes local variations affecting in particular the peak emissivity obtained from the tomographic inversion. Foil non-uniformities are planned to be reduced in the future by replacing the foil with one produced with vapour deposition processes\cite{Federici2024}. Additionally, the radiation front is not strongly localised on MAST-U as both impurity radiation and, downstream of the impurity radiation region, electron-impact excitation (EIE) results in significant radiative power loss \cite{Verhaegh2023b}. EIE is the largest contributor to the total radiated power \emph{within} the \emph{divertor chamber} (in a purely hydrogenic plasma\cite{Verhaegh2023a,Verhaegh2023b}). Due to the size of the radiation region and foil non-uniformities, the radiation peak location cannot be obtained reliably. However, the location where the radiation has a sharp decrease can be tracked more reliably and is likely reminiscent of a thermal front where $T_e$ becomes too low ($<3-5$ eV) for EIE and ionisation \cite{Verhaegh2023a}. Hence, the radiation region or 'front' is tracked experimentally also by tracking the location where the radiative emissivity reaches a set fraction (50\% is used here) of the peak emission in the divertor leg, analogous to techniques that have tracked the ionisation front using imaging \cite{Wijkamp2022}.

The peak radiative emissivity corresponds to the peak in total radiation (hydrogenic + impurity), whose movement we define as the beginning of 'radiative detachment'. This is not to be confused with (particle) detachment, when the ionisation source detaches from the target and the ion target flux rolls-over, which is in better agreement with the 50\% falloff point of the radiative emissivity.
In the remainder of the paper, both the peak of radiative emissivity and the 50\% falloff are tracked in terms of poloidal distance from the target along the separatrix divided by the poloidal distance from the X-point to target, called ${\hat{L}}_{x}$: 0 corresponds to the target, 1 corresponds to the X-point, $>1$ implies locations upstream of the X-point towards the midplane. It should be noted that given the IRVB FOV, the radiation cannot be reliably tracked on the outer separatrix past the X-point.


To put the evolution of these two radiative power markers into perspective, their evolution is compared with the total target particle flux. Before detachment, the target particle flux first increases as the upstream/core density is decreased, as predicted by the 2PM. Due to a combination of power limitation, volumetric recombination and momentum losses, the particle flux first plateaus (detachment onset) and then decreases or 'rolls-over' (detachment). In MAST-U, Langmuir probes (LPs) are used to monitor the particle flux, and they are present only in the outer strike points (both upper and lower in DN).\cite{Ryan2023} The roll-over of the particle flux is expected to be associated with the detachment of the ionisation region, and thus (if hydrogenic radiation is at least comparable with the impurity driven one) the 50\% fall-off point of the radiative emissivity, from the target. The detachment of the ionisation region from the target can be observed more directly using the $D_2$ Fulcher band emission front as a proxy for the ionisation region \cite{Wijkamp2022,Verhaegh2022}. For a CD, this can be observed by monitoring the brightness of the $D_2$ Fulcher emission near the strike point using the DMS diagnostic \cite{Verhaegh2022}. Although not shown here, the reduction of $D_2$ Fulcher brightness at the target is in agreement with the ion target flux roll-over within experimentally uncertainties (e.g. it occurs at slightly higher upstream density ($0.05/0.1 \cdot 10^{19} \#/m^3$)). This confirms the self-consistency of these two detachment metrics.

Finally, a movement of the radiation front upstream implies that a region of 'cold' plasma moves towards the core. Work on other tokamaks has shown that core confinement deteriorates when this radiation front is near the X-point and, subsequently, enters the core \cite{Kallenbach2015a,Reinke2013,Reimold2015}. Such core deterioration is detrimental for future fusion reactors. To monitor this, we compare the evolution of detachment with the energy confinement time, ${\tau }_{ th }$, defined by $\frac {dW} {dt}={P}_{heat} - \frac {W} {{\tau }_{ th }}$ with $W$ the stored energy and ${P}_{heat}$ the power injected in the core.

\begin{figure*}
	\centering
	\includegraphics[trim={150 30 180 80},clip,width=0.9\linewidth]{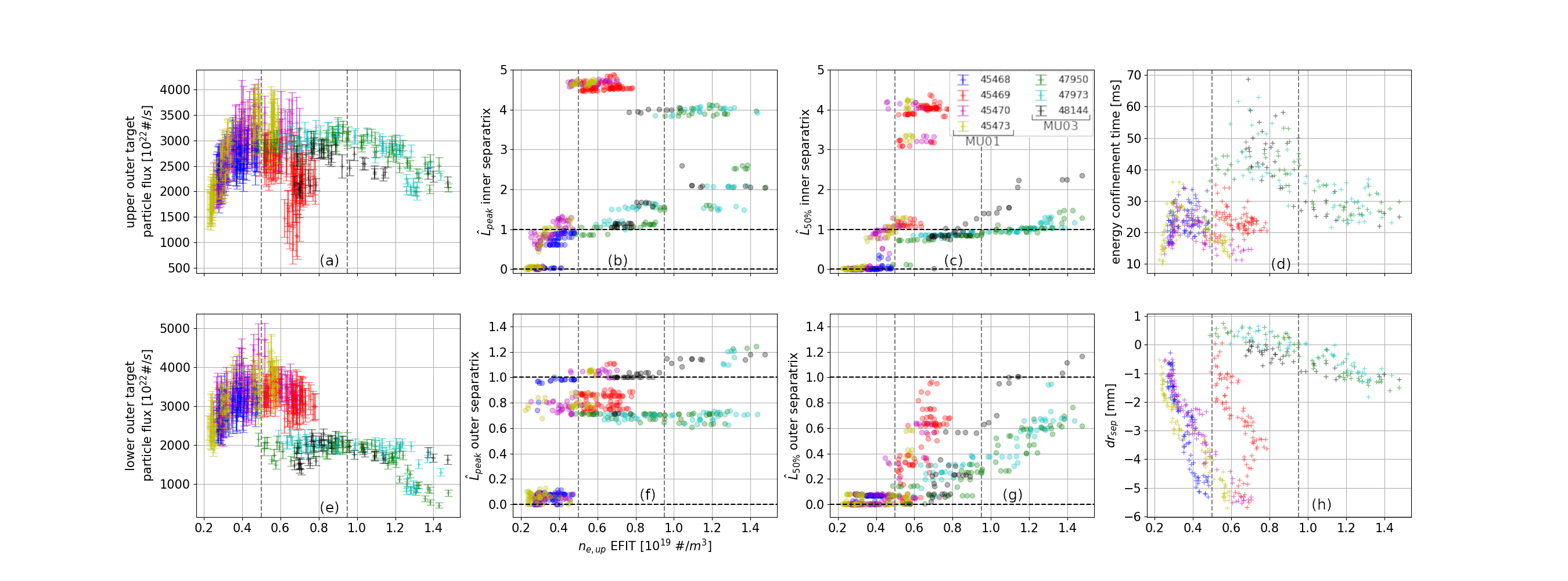}
         \vspace{-4mm}
	\caption{(a) upper and (e) lower target current, providing the detachment condition from particle flux roll-over compared to the upstream density determined using Thomson scattering (TS) and EFIT. (b) movement of the radiated emission peak and (c) of the region with emissivity of 50\% of the peak on the separatrix, both indicating radiative detachment, in the inner leg and (f, g) lower leg. (d) energy confinement time and $dr_{sep}$ (h) variation with upstream density. Data referring to Ohmic heated L-mode shots. The first 4 cases are from MU01 with HFS fuelling, Ip=600kA and $P_{SOL}\sim$0.4MW while last 3 are from MU03 with LFS fueling, $I_p$=750kA and $P_{SOL}\sim$0.5-0.6MW. The vertical lines indicate the approximate particle flux roll-over.}
	\label{fig:ohmic}
    \vspace*{-7mm}
\end{figure*}

\section{Front tracking results}\label{Front tracking results}

After having discussed the general evolution of the radiative emissivity and defined our two different radiative front identifiers, we show their evolution as function of upstream density at the outer separatrix midplane ($n_{e,up} EFIT$). $n_{e,up} EFIT$ is obtained from a smoothed Thomson scattering (TS) data and the location of outer midplane separatrix from EFIT\cite{Lao1985}. An important caveat to this technique is that $n_{e,up} EFIT$ has large uncertainties due to the uncertainty of the separatrix position determination. Although this can cause systematic uncertainties between different discharges, the trend here identified by $n_{e,up} EFIT$ should be relatively reliable, as similar dependencies are found in terms of Greenwald fraction. The density on the inner side is supposed to be the same, as the variation of the plasma parameters on the same flux surface in the core is limited, although this is not necessarily true for HFS fuelled discharges\cite{Moulton2024}.

\autoref{fig:ohmic} compares our tracking results with the ion target flux measurements, energy confinement time and $dr_{sep}$ for the MU01 and MU03 Ohmic shots. 

In MU01, the roll-over is quite clear on both outer strike points, happening at $n_{e,up} EFIT \sim 0.5 \times 10^{19} \#/m^3$, and the particle flux is quite up/down symmetric. This is the case even if in these shots $dr_{sep}$ (the distance between the lower and upper separatrix at the outer midplane; if positive the core is more strongly coupled with the upper divertor and vice versa) is 0/-6mm, with $\lambda_q$ in the range 5-15 mm (consistent with MAST scaling laws \cite{Harrison2013}).Such a large $dr_{sep}/\lambda_q$ ratio is expected to cause significant heat flux asymmetries  \cite{Fevrier2021}, which seems to be inconsistent with the particle flux symmetry and the lack of asymmetries observed in high speed camera data or resistive bolometry measurements. This may indicate inaccuracies in the $dr_{sep}$ retrieved from EFIT for MAST-U or larger $\lambda_q$ than currently determined or expected from scaling laws. 

The start of the peak radiation movement from the inner target across all MU01 discharges happens at a slightly lower $n_{e,up}$ than the outer peak radiation movement and thus at a lower $n_{e,up}$ than the particle flux roll-over. The movement of the peak radiation is not regular for both divertor legs and there appear to be discrete steps, likely due to the foil non uniformity previously mentioned. However, the radiation evolution difference is more noticeable on the 50\% falloff marker. As mentioned, because these discharges are fuelled from the HFS, the radiation on the inner separatrix reaches the HFS midplane before the outer leg is fully detached with both markers. The evolution of the 50\% falloff marker is very gradual on the outer divertor leg, in contrast to the inner leg where the radiation quickly disappears from the target and re-appears near the X-point (e.g. there are almost no points between $\hat{L}=$0 and 0.8). This abrupt movement of the radiation on the inner leg, but not on the outer leg, is in agreement with the DLS model prediction in \autoref{Model predictions}, of a stable detachment on the outer leg and unstable detachment on the inner leg. Further research is required using post MU02 discharges where the IRVB viewing geometry was fully verified to obtain further confidence in this conclusion. 
Lastly, the confinement time initially increases due to the increase in stored energy at the beginning of the shot, but as soon as the peak emission moves from the target there is a strong degradation. This is true also for the ratio of ${\tau }_{ th }$ with the confinement time from scaling laws identified in \cite{Federici2023b}. This is most likely due to the negative effect of regions with high emissivity and low temperature inside or close to the core.

Ohmic shots from MU03 have higher $I_p$ and $P_{SOL}$ and reduced interactions with the main vessel and baffle plates. This increases the particle flux roll-over point from a Greenwald fraction of 0.22 to 0.35. Although $\mid dr_{sep} \mid /\lambda_q$ is reduced (e.g. $\mid dr_{sep} \mid <1mm$ with $\lambda_q$ between 5-15 mm), the LPs measure a noticeably higher particle flux on the
upper outer leg. This is consistent with recent experiments and simulation studies that indicate significant up/down asymmetries in the MAST-U plasma due to drifts in a connected DN, although further studies are required - particularly given the uncertainty of $dr_{sep}$.\cite{Lovell2024,ParadelaPerez2024} The initial upstream density is not not low enough to witness the detachment of the peak emission on either leg, but the 50\% falloff still detaches from the inner target at a lower $n_{e,up}$ than at the outer target, much earlier than the outer target roll-over. A rapid movement of the inner leg radiation region towards the X-point is still observed, although only a few points of $\hat{L}_{50\%} \approx 0$ are observed. The outer leg $\hat{L}_{50\%}$ evolution is more gradual, in agreement with the older MU01 results. As the density increases, the emissivity on the entire inner separatrix increases, with similar peak emissivities near the midplane and the X-point (causing $\hat{L_{peak}}$ to oscillate), whereas  the 50\% falloff marker remains quite stable near the X-point. When the outer leg radiation reaches the X-point, the radiation on the inner separatrix rises further upstream. Apart from the very end of shot 47950, there is no evidence of the presence of a MARFE-like structure localised at the HFS midplane from both IRVB and resistive bolometry. The lack of such a MARFE structure results in less degradation of $P_{SOL}$, likely explaining why the outer target flux roll-over and all phases of radiative detachment occur over a much larger range of upstream density. The energy confinement time is increased and seems to peak at a higher $n_{e,up}$ than the start of the outer leg radiative detachment (${\hat{L}}_{peak}>0$) and the inner separatrix radiation reaching the X-point ($\hat{L_{50\%}} \approx 1$). The relative decrease of confinement time is also lower than in MU01 and occurs at much higher $n_{e,up}$.However, particle detachment (e.g. the outer target particle flux roll-over) happens at a higher $n_{e,up}$ than the confinement peak; implying that outer target detachment (for the CD), without impurity seeding, requires a degradation of confinement for this Ohmic L-mode scenario.
\begin{figure*}
	\centering
	\includegraphics[trim={150 30 180 80},clip,width=0.9\linewidth]{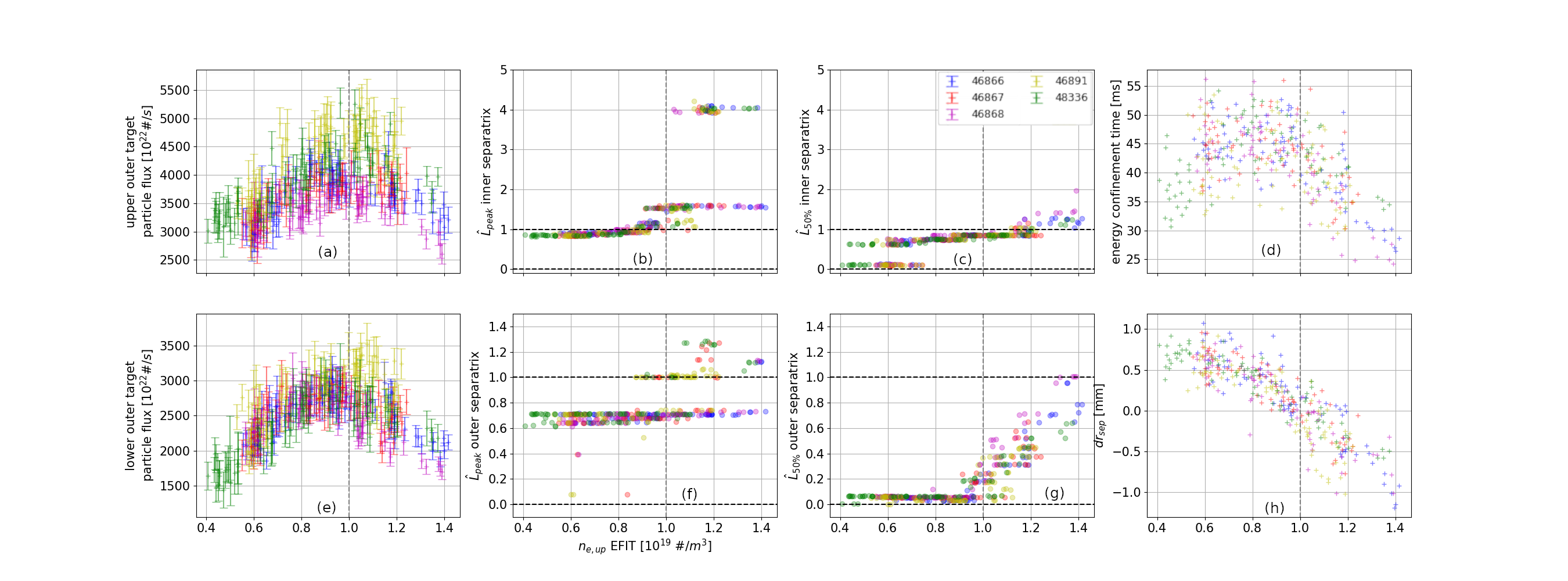}
         \vspace{-4mm}
	\caption{The quantities are the same as in \autoref{fig:ohmic}. Data referring to beam heated L-mode shots ($P_{SOL} \sim 1-1.5MW$) fuelled from the LFS and $I_p$=750kA.}
	\label{fig:beamHeated}
    \vspace*{-7mm}
\end{figure*}

The results for the MU03 beam heated L-mode shots are shown in \autoref{fig:beamHeated}. The ratio of up/down outer particle flux is unchanged, but both particle fluxes are increased due to the higher $P_{SOL}$. The outer target ion flux increases initially more sharply than observed in the Ohmic cases, likely as the initial density achieved is lower and $P_{SOL}$ is significantly higher. Because of this, the ion target flux evolution shows a clearer roll-over - although the reduction of ion target flux occurs at similar core and upstream densities. Due to the lower $n_{e,up}$/higher $P_{SOL}$ starting point, it is possible to observe the inner leg $\hat{L_{50\%}}$ unstable transition from the target to the X-point, consistent with the DLS model predicion. Curiously, $\hat{L_{50\%}}$ transitions from 0 to 0.6, rather than 0.8. This could be consistent with a larger extent of the radiation front. Including the extent of the thermal front (rather than assuming it is infinitely narrow) is one of the goals of the DLS-Extended model\cite{Kryjak2024}. A larger front has the effect of averaging the magnetic characteristics in a set region of the leg, providing a larger stability window for the front. The outer leg 50\% falloff marker movement happens at about the same density as the particle flux roll-over, at much higher $n_{e,up}$ compared to the inner leg. 
The energy confinement time profile is quite similar to the MU03 Ohmic case, suggesting that the increased external heating results in a proportional stored energy increase. 
The confinement peak occurs while $\hat{L_{50\%}}$ approaches 1 on the inner leg, but before it starts to increase on the outer leg. This seems to imply that the difference in outer leg detachment does not have a significant impact on confinement. This might be due solely to the inner leg detaching first and the radiation starting to cross inside the core from there, and if a different partition of $P_{SOL}$ between inner and outer leg could be achieved the relationship could be reversed. It might be possible to achieve this in single null, when more power is directed to the inner leg.

\section{Conclusion and future work}\label{Conclusion and future work}

In this paper the first scientific results exploiting the new MAST-U infrared imaging bolometer (IRVB) are presented using L-mode conventional divertor density ramp discharges. The radiation along the inner divertor leg sharply transitions from near the target to near the X-point as density is increased. In contrast, the radiation front detachment evolves gradually from the target to the X-point. The DLS model \cite{Lipschultz2016,Cowley2022} suggests an unstable thermal front evolution at the inner target and a stable thermal front evolution at the outer target, consistent with experiments. After both radiation fronts have detached from the target, ultimately particle detachment is observed from the ion target flux roll-over.

The fueling location can have a strong impact on the radiation evolution: fueling from the HFS causes the emergence of a MARFE-like structure at the HFS midplane that can then penetrate the core and effect core performance. Resolving this limitation and reducing plasma surface interactions with the main vessel wall resulted in a clearer separation between radiation detachment and particle flux detachment on the outer leg. The degradation of the energy confinement time occurs at higher upstream densities than the complete radiative detachment on the inner leg, but lower than the particle flux detachment.

Hardware upgrades for improved measurements are underway, including a replacement of the existing IRVB IR camera,  improving signal to noise levels and enabling a higher time resolutions, to improve the monitoring of transients. In the next MAST-U vacuum breach, the foil will be replaced with a more uniform one\cite{Federici2024}, and a second IRVB installed aimed at the upper X-point, to assess up/down asymmetries and provide a full bolometric coverage of the entire plasma volume. Future experimental campaigns, that should try to start from a lower initial upstream density, are planned to investigate higher power H-mode conditions, up/down symmetries\cite{Lovell2024,ParadelaPerez2024} and the impact of inner target geometry.

\begin{acknowledgments}
\fontsize{7.5pt}{8pt}\selectfont
This work is supported by US Department of Energy, Office of Fusion Energy Sciences under the Spherical Tokamak program, contract DE-AC05-00OR22725 and under the auspices of the EPSRC [EP/L01663X/1]. Support for M. L. Reinke’s contributions was in part provided by Commonwealth Fusion Systems.

This work has been carried out within the framework of the EUROfusion Consortium, funded by the European Union via the Euratom Research and Training Programme (Grant Agreement No 101052200-EUROfusion) and from the EPSRC [grant number EP/W006839/1]. Views and opinions expressed are however those of the author(s) only and do not necessarily reflect those of the European Union or the European Commission. Neither the European Union nor the European Commission can be held responsible for them.


\end{acknowledgments}

\section{References}
  \vspace{-3mm}

\begin{thebibliography}{0}%
\makeatletter
\providecommand \@ifxundefined [1]{%
 \@ifx{#1\undefined}
}%
\providecommand \@ifnum [1]{%
 \ifnum #1\expandafter \@firstoftwo
 \else \expandafter \@secondoftwo
 \fi
}%
\providecommand \@ifx [1]{%
 \ifx #1\expandafter \@firstoftwo
 \else \expandafter \@secondoftwo
 \fi
}%
\providecommand \natexlab [1]{#1}%
\providecommand \enquote  [1]{``#1''}%
\providecommand \bibnamefont  [1]{#1}%
\providecommand \bibfnamefont [1]{#1}%
\providecommand \citenamefont [1]{#1}%
\providecommand \href@noop [0]{\@secondoftwo}%
\providecommand \href [0]{\begingroup \@sanitize@url \@href}%
\providecommand \@href[1]{\@@startlink{#1}\@@href}%
\providecommand \@@href[1]{\endgroup#1\@@endlink}%
\providecommand \@sanitize@url [0]{\catcode `\\12\catcode `\$12\catcode
  `\&12\catcode `\#12\catcode `\^12\catcode `\_12\catcode `\%12\relax}%
\providecommand \@@startlink[1]{}%
\providecommand \@@endlink[0]{}%
\providecommand \url  [0]{\begingroup\@sanitize@url \@url }%
\providecommand \@url [1]{\endgroup\@href {#1}{\urlprefix }}%
\providecommand \urlprefix  [0]{URL }%
\providecommand \Eprint [0]{\href }%
\providecommand \doibase [0]{http://dx.doi.org/}%
\providecommand \selectlanguage [0]{\@gobble}%
\providecommand \bibinfo  [0]{\@secondoftwo}%
\providecommand \bibfield  [0]{\@secondoftwo}%
\providecommand \translation [1]{[#1]}%
\providecommand \BibitemOpen [0]{}%
\providecommand \bibitemStop [0]{}%
\providecommand \bibitemNoStop [0]{.\EOS\space}%
\providecommand \EOS [0]{\spacefactor3000\relax}%
\providecommand \BibitemShut  [1]{\csname bibitem#1\endcsname}%
\let\auto@bib@innerbib\@empty
\end{thebibliography}%


\begin{thebibliography}{9}
\fontsize{7.5pt}{8pt}\selectfont

\bibitem{Federici2023}
F.~Federici~et al., \emph{Review of Scientific Instruments}, vol.~94, no.~3, p.
  033502, 3 2023.

\bibitem{Federici2023a}
F.~Federici~et al, in \emph{70th Annual Meeting of the APS Division of Fluid
  Dynamics}, Denver, 2023, p.~1.

\bibitem{Myatra2021}
O.~Myatra, Ph.D. dissertation, University of York, 2021.

\bibitem{Cowley2022}
C.~Cowley~et al., \emph{Nuclear Fusion}, vol.~62, no.~8, 2022.

\bibitem{Lipschultz2016}
B.~Lipschultz~et al., \emph{Nuclear Fusion}, vol.~56, no.~5, 2016.

\bibitem{Morris2018}
W.~Morris~et al., \emph{IEEE Transactions on Plasma Science}, vol.~46, no.~5,
  pp. 1217--1226, 2018.

\bibitem{Fishpool2013}
G.~Fishpool~et al., \emph{Journal of Nuclear Materials}, vol. 438, no. SUPPL,
  pp. S356--S359, 7 2013.

\bibitem{Moulton2024}
D.~Moulton~et al., \emph{Nuclear Fusion}, pp. 1--50, 5 2024.

\bibitem{Verhaegh2022}
K.~H. Verhaegh~et al., \emph{Nuclear Fusion}, vol.~63, no.~1, p.~21, 2022.

\bibitem{Verhaegh2023a}
K.~Verhaegh~et al., \emph{Nuclear Fusion}, vol.~63, no.~12, 2023.

\bibitem{Soukhanovskii2022a}
V.~A. Soukhanovskii~et al., \emph{Nuclear Materials and Energy}, vol.~33, no.
  June, 2022.

\bibitem{Henderson2024}
S.~S. Henderson~et al., \emph{Submitted to Nuclear Materials and Energy}, 2024.

\bibitem{Federici2023b}
F.~Federici, Ph.D. dissertation, University of York, 2023.

\bibitem{Federici2024}
F.~Federici~et al., \emph{Submitted to Review of Scientific Instruments}, pp.
  1--5, 2024.

\bibitem{Silburn2020}
S.~A. Silburn~et al., ``{Calcam},'' 2020.

\bibitem{Rivero-Rodriguez2018}
J.~F. Rivero-Rodriguez~et al., in \emph{45th EPS Conference on Plasma Physics,
  EPS 2018}, vol. 2018-July.\hskip 1em plus 0.5em minus 0.4em\relax European
  Physical Society, 2018, pp. 233--236.

\bibitem{Verhaegh2023b}
K.~Verhaegh~et al., \emph{submitted to Nuclear Fusion}, pp. 1--30, 11 2023.

\bibitem{Stangeby2001}
P.~Stangeby, ser. Series in Plasma Physics.\hskip 1em plus 0.5em minus
  0.4em\relax CRC Press, 1 2000, vol.~43, no.~2.

\bibitem{Hutchinson1994}
I.~H. Hutchinson, \emph{Nuclear Fusion}, vol.~34, no.~10, pp. 1337--1348, 1994.

\bibitem{Harrison2013}
J.~Harrison, G.~Fishpool, and A.~Kirk, \emph{Journal of Nuclear Materials},
  vol. 438, pp. S375--S378, 2013, proceedings of the 20th International
  Conference on Plasma-Surface Interactions in Controlled Fusion Devices.

\bibitem{Wijkamp2022}
T.~A. Wijkamp~et al., \emph{Nuclear Fusion}, vol.~63, no.~5, p. 056003, 5 2023.

\bibitem{Ryan2023}
P.~J. Ryan~et al., \emph{Review of Scientific Instruments}, no. to be
  submitted, 2023.

\bibitem{Kallenbach2015a}
A.~Kallenbach~et al., \emph{Nuclear Fusion}, vol.~55, no.~5, 2015.

\bibitem{Reinke2013}
M.~L. Reinke, in \emph{7th IAEA Technical Meeting on Steady State Operation of
  Magnetic Fusion Devices}, 2013.

\bibitem{Reimold2015}
F.~Reimold~et al., \emph{Nuclear Fusion}, vol.~55, no.~3, 2015.

\bibitem{Lao1985}
L.~L. Lao~et al., \emph{Nuclear Fusion}, vol.~25, no.~11, pp. 1611--1622, 1985.

\bibitem{Fevrier2021}
O.~F{\'{e}}vrier~et al., \emph{Nuclear Fusion}, vol.~61, no.~11, 2021.

\bibitem{Lovell2024}
J.~J. Lovell~et al., \emph{Submitted to Nuclear Materials and Energy}, 2024.

\bibitem{ParadelaPerez2024}
I.~Paradela~P{\'{e}}rez, \emph{Nuclear Fusion, in preparation}, 2024.

\bibitem{Kryjak2024}
M.~Kryjak, 50th annual IOP Plasma Physics Conference, York, Tech. Rep., 2024.

\end{thebibliography}

\end{document}